\documentclass[sigconf]{acmart}

\usepackage{booktabs} 

\setcopyright{rightsretained}

\acmDOI{10.475/123_4}

\acmISBN{123-4567-24-567/08/06}

\acmConference[WI-IAT 2017]{IEEE/WIC/ACM International Conference on Web Intelligence 2017}{August 2017}{Leipzig, Germany} 

\begin{document}
\title{Turned 70? It is time to start editing Wikipedia.}

\author{Radoslaw Nielek}
\orcid{0000-0002-5794-7532}
\affiliation{%
  \institution{Polish-Japanese Academy of Information Technology}
  \streetaddress{ul. Koszykowa 86}
  \city{02-008 Warsaw} 
  \state{Poland} 
}
\email{nielek@pja.edu.pl}

\author{Marta Lutosta\'{n}ska}
\affiliation{%
  \institution{Senfino Inc.}
  \streetaddress{Wawozowa 11}
  \city{02-796 Warsaw} 
  \state{Poland} 
}
\email{marta.lutostanska@pja.edu.pl}

\author{Wies\l{}aw Kope\'{c}}
\affiliation{%
  \institution{Polish-Japanese Academy of Information Technology}
  \streetaddress{ul. Koszykowa 86}
  \city{02-008 Warsaw} 
  \state{Poland} 
}
\email{kopec@pja.edu.pl}

\author{Adam Wierzbicki}
\affiliation{%
  \institution{Polish-Japanese Academy of Information Technology}
  \streetaddress{ul. Koszykowa 86}
  \city{02-008 Warsaw} 
  \state{Poland} 
}
\email{adamw@pja.edu.pl}

\renewcommand{\shortauthors}{R. Nielek et al.}

\begin{abstract}
Success of Wikipedia would not be possible without the contributions of millions of anonymous Internet users who edit articles, correct mistakes, add links or pictures. At the same time Wikipedia editors are currently overworked and there is always more tasks waiting to be completed than people willing to volunteer. The paper explores the possibility of involving the elderly in the Wikipedia editing process. Older adults were asked to complete various tasks on Wikipedia. Based on the observations made during these activities as well as in-depth interviews, a list of recommendation has been crafted. It turned out that older adults are willing to contribute to Wikiepdia but substantial changes have to be made in the Wikipedia editor.
\end{abstract}

%
%

\ccsdesc[500]{Human-centered computing~Wikis}
\ccsdesc[300]{Human-centered computing~Usability testing}
\ccsdesc[300]{Human-centered computing~Empirical studies in interaction design}
\ccsdesc[300]{Human-centered computing~Empirical studies in collaborative and social computing}

\keywords{Wiki, Wikipedia, older adults, participatory design, content creation}

\maketitle

\section{Introduction}

Success of Wikipedia would not be possible without contributions of millions of anonymous Internet users who edit articles, correct mistakes, add links or pictures. Future development of the Wikipedia is endangered by decreasing number of active editors. Those who contribute to Wikipedia are overworked and Wikipedia faces the competition for the most skilled and passionated volunteers from other crowdsourcing projects (e.g. citizen sciences platforms like Zooniverse\footnote{\url{https://zooniverse.org}}). There are always more tasks waiting to be completed than people who are willing to volunteer and devote their spare time.

At the same time societies are getting older and there are more and more retired people who are in a good physical and cognitive shape. They look for meaningful and enjoying activities, which help them feel needed and maintain their self-esteem. Contributing to Wikipedia seems to be a good task for that purpose as it requires balanced opinion and good language skills, which are more common among active older adults. Moreover, editing Wikipedia by older adults can be mutually beneficial -- for the elderly and the community. Such bidirectional benefits of engagement in online content production communities was already shown for students \cite{Farzan:2013:WCE:2470654.2470765,roth2013assigning} but to the best of authors' knowledge no such intervention was tried on older adults.

This study was crafted to verify the hypothesis that older adults may be a good Wikipedia editors and to identify barriers that might limit their engagement. Ten older adults with the average age of 69.3 were asked to complete seven tasks on Wikipedia (among others: creating account, adding picture, correcting misspellings etc.). Participants were observed by a researcher while performing the tasks. The study was followed up by the affinity group interview. 

Results collected during the study show that long-term goal of increasing percentage of older adults among Wikipedia editors seems doable but it requires some changes in the Wikipedia editor and editing process. More cohesion in the system, more similarities with word processors, and less cognitively demanding editor are among the most important ones. Although the whole study was conducted on Polish Wikipedia most of observations can be easily and directly generalized not only to another language versions but also to the other Wikis.

The remaininig part of the paper is organized as follow. Next section summarizes related works. Research methodology is described in third section. Fourth section presents results and is followed by discussion in the fifth section. The last section concludes the paper.

\section{Related works}

Contribution of older adults to online communities and participation in crowdsourcing projects have been gaining more and more attention from researchers in recent years. Brake \cite{brake2014we} addressed the question whether need for creating digital content and sharing it on the web is universal among people. Japanese researchers developed a crowdsourcing platform dedicated to engaging older adults in proofreading of scanned books \cite{itoko2014involving,kobayashi2013age}. Moreover, Kobayashi et al. \cite{kobayashi2015motivating} used the same system to study different strategies for motivating older adults. To overcome technological barriers and lack of motivation Hiyama et al. \cite{hiyama2013question} proposed a passive interaction model for acquiring knowledge from seniors. Recent study conducted by Brewer et al.\cite{brewer2016would} shows that older adults have neither required skills nor motivation to work on the most popular microtask platform -- Amazon Mechanical Turk.

Pan et al. \cite{pan2015effects} have identified three types of design familiarity that are crucial to assure adoption by older adults: (a) \textit{symbolic familiarity}, (b) \textit{cultural familiarity}, and (c) \textit{actionable familiarity}. Researchers from The Joint NTU-UBC Research Centre of Excellence in Active Living for the Elderly in Singapore\footnote{http://www.ntulily.org/} tried to use this approach to develop a mobile application for tagging historical photos by retired Singaporeans \cite{yu2016productive}. An overview of tools for online contribution of older adults is presented in \cite{ibarratools}.

Wikipedia as the most successful example of crowdsourcing system has been thoroughly studied. Researchers tries to predict articles' credibility \cite{jankowski2014predicting} or reveal the patterns of collaboration among editors and learn from them \cite{turek2010learning,turek2011wikiteams,wierzbicki2010learning} but, to the best of authors' knowledge, no research has been done in the subject of older adults as Wikipedia editors and contributors. Vora et al. \cite{vora2010n00b} conducted qualitative study to identify and reduce technical barriers encountered by first-time editors. Some attempts have also been made to improve accessibility of Wikipedia editor tool for the visually impaired people \cite{buzzi2008making} and to implement WAI-ARIA\footnote{Accessible Rich Internet Applications Guidelines -- \url{https://www.w3.org/TR/wai-aria/}} guidelines  \cite{senette2009enhancing}. Narayan et al. \cite{narayan2015effects} proposed the Wikipedia Adventure game designed for familiarizing new users with norms, practices and policies existing on Wikipedia. Morgan et al. \cite{morgan2013tea} designed a platform called "Teahouse" for mixing new Wikipedia editors with experienced members of the community to share tips and help solving encountered problems. Ciampaglia and Taraborelli \cite{ciampaglia2015moodbar} noticed that socialization tools may cause a workload increase for experienced Wikipedian and proposed a "lightweight socialization tool" for new users.

In theory everyone can contribute to Wikipedia, but in reality only a very small number of people do it regularly and they strongly influence the overall tone of Wikipedia articles \cite{lee2016crowdsourcing}. On the other hand, the remaining majority is crucial for correcting mistakes and fine-tuning articles \cite{halfaker2013making}. Therefore, encouraging and motivating episodic editors is important.    

As virtually all tasks on Wikipedia are carried out by volunteers, motivation is crucial for long-term sustainability of the on-line encyclopedia. Yang and Cheng-Yu \cite{yang2010motivations} showed that "internal self-concept motivation" is the main driving force behind Wikipedia editors. Based on a survey of 100 university students Staccy Kuznetsov \cite{Kuznetsov:2006:MCW:1215942.1215943} mentions five possible motives: altruism, reciprocity, community, reputation, and autonomy. Qualitative insight into the external perception of Wikipedia editors provides work published by Judd Antin \cite{antin2011my}.   

Participation in volunteering activity has many positive effects on the elderly. \cite{lum2005effects} claim that older adults who frequently volunteer tend to have improved physical and mental health, compared to those who do not participate in volunteering. The work of \cite{morrow2003effects} extends this positive correlation to well-being. Effects of formal volunteering are also presented by \cite{greenfield2004formal}, where authors claim that such activities serve as a protective factor for the older adults' psychological well-being. Such observation is also rectified by \cite{hao2008productive}. This is crucial, since studies also show that in some regards, elders are more likely to be engaged in volunteering activity \cite{morrow2010volunteering}. Despite some early studies \cite{dickinson2006computer} researchers agree that using computer and Internet improves well-being of older adults either directly \cite{Seiderer:2015:EDI:2750511.2750514} or indirectly \cite{Brewer:2016:TLR:2858036.2858379}.

Finally, although there is no research studying differences between young and older adults with respect to editing Wikipedia, some researchers have addressed sex differences. Only a dozen or so percent of Wikipedia editors are women \cite{antin2011gender}. Lam et al. \cite{lam2011wp} tried to explain this bias by conflict-related factors and figuratively compared Wikipedia to clubhouse. Some results confirm that gender bias in crowdsourcing projects understood as lower participation and contribution rate is common among different crowdsourcing sites \cite{chang2014specialization,forte2012some}. 

\section{Experimental setup}

\subsection{Methodology}

Selecting methods and setup for our study, we followed the recommendation crafted by Dickinson et al. \cite{dickinson2007methods} for maximizing the research outcomes of working with older adults.

We decided to use a mixed approach by employing direct observation methods and results of participants' learning process. In particular, the study is comprised of three major phases: individual on-site activity, individual remote site activity and final on-site common wrap-up workshop.

In the first phase we employed a set of complimentary qualitative methods and tools i.e. IDI, usability testing and thinking aloud protocol. Each individual session was started by an in-depth interview to explore the context of the participant's experience and then followed by the usability tests performed along the predefined scheme with the use of the thinking aloud method to enhance the moderator's observation. Each session lasted less than an hour. Participant's voice and face image was recorded alongside with corresponding full image of the computer screen with the use of the Camtasia Studio software.

Interviews were precluded by short questionnaire with questions about age, education level and acquired digital skills (i.e. frequency of using the Internet). 

In the next phase participants were proposed to continue their off-site, i.e. work at home.  After a week there was a final phase with a workshop session organized for all participants to wrap-up the study and facilitate joint discussion about their experiences and remarks.

\subsection{Participants}

Participants were recruited from a local senior club. The main motivation for them was to learn a little bit about Wikipedia, thus, we did not offer any additional incentives. 10 seniors took part in individual interviews (7 woman and 3 man). The average age was 69.3, SD=8.08. The oldest participant was 87 years old (the youngest one 62). All of them live in Warsaw (capital of Poland) and are socially very active.

According to the self-declared data from questionnaire that precluded interviews, older adults taking part in our study were well educated (majority with university degree) and all of them use Internet on regular basis (e-mail and on-line banking are among the most popular services) on their own laptops (8 people) or desktops (2 people). Six of ten participants have Facebook accounts.

We have deliberately chosen users who stand out from the background of their age group with above average computer skills, education and literacy. Editing Wikipedia is a complex task that requires not only familiarity with technology but also broad knowledge, ability to analyze and process information as well as writing skills. Therefore, it is reasonable to assume that only a small fraction of older adults may be considered as valuable editors.

\subsection{Tasks}
Participants were confronted with seven tasks. The next task was given to the participant only after the completion of the the current one. All tasks were related to either two, relatively noncontroversial and well-known to Poles, historic events -- "Battle of Warsaw"\footnote{\url{https://en.wikipedia.org/wiki/Battle_of_Warsaw_(1920)}} and "Battle of Vienna"\footnote{\url{https://en.wikipedia.org/wiki/Battle_of_Vienna}} or a less known but also uncontroversial term "literacy cafe"\footnote{Cafes in Warsaw during interwar period where at that time gathered famous poets and writers}. Brief description of all tasks is given below:

\begin{enumerate}
\itemsep0em 
\item Imagine that you just watched on TV some news about celebration of the anniversary of Battle of Warsaw. Please, check in the Wikipedia how this battle is described. Take a closer look on the section dedicated to the importance of the Battle of Warsaw for Europe.
\item It turned out that there is an error in the article about "Battle of Warsaw". Twice incorrectly introduced 1922 instead of 1920. Please, try to correct this error.
\item In the article about "Battle of Warsaw" another important historical event is mentioned -- "Battle of Vienna" -- but there is no direct link to this article. Please, add it. 
\item In the twenties of the last century, literary cafes were popular in Warsaw but the article describing it does not exist on the Wikipedia. Please, try to add this article using our prepared text that you will find in the folder on the desktop.
\item Do you want to further support the Wikipedia by editing articles? Shall we try to create account? 
\end{enumerate}

\section{Results}

\subsection{Wikipedia in the eyes of seniors}

\subsubsection{Wikipedia contra traditional encyclopedia}

It came out that all participants used Wikipedia on regular basis for checking information and new terms. Moreover, most older adults in our study treated Wikipedia as an important part of their world -- 
\textit{"If it wasn't possible to check in Wikipedia I would sometimes feel uncomfortable""} (M68). When asked to compare Wikipedia with the traditional Encyclopedia participants pointed at \textit{"broader range of knowledge"}, \textit{"more content"} and \textit{"more elaborate articles"}. 

\begin{quote}
\textit{The paper one is definitively smaller. Although, it's still over twenty volumes, (...) But nobody wants to use it now. I'd say that it's like and old telephone -- it's something that almost nobody uses it.} (F77)
\end{quote}

Opinions expressed by another participant not only support this view but also reveal that he notice the advantages of the Wikipedia. The Wikipedia is perceived as a "better encyclopedia".

\begin{quote}
\textit{"In Wikipedia it's easy with those links -- you can go in one entry and find everything related to it, it's easier to find additional explanations with key words."} (M87)
\end{quote}

\subsubsection{Credibility of entries}

Less enthusiastic opinions were expressed when participants were asked about the credibility of the Wikipedia articles. Although they admitted that they had never encountered a misstep or error in Wikipedia, their distrust was raised by a variety of heard opinions.

\begin{quote}
\textit{"Everybody around me says that is unreliable, because it isn't created by very competent people. Sometimes information is put there by people that don't know much, but so far I haven't encountered any substantial errors. If I found something, I though rather it was good."} (F64)
\end{quote}

\begin{quote}
\textit{"I'd give 95\% for yes, but leave 5\% for no -- people aren't sometimes honest on the Internet -- paper absorbs all."} (F67) 
\end{quote}

None of the participants was able to give any specific example of either non-credible information in Wikipedia or sources of their concerns. They used terms like: \textit{"I heard"}, \textit{"somebody told me"} or \textit{"everyone is talking"}.

\begin{quote}
\textit{"I've heard that everybody may write some entry. I think that if somebody doesn't have the appropriate knowledge, it won't work. I think that such access shouldn't be given to everybody, because there are some smartasses who think that they know everything."} (F77)
\end{quote}
\begin{quote}
\textit{"I don't actually trust it. (...) Those may be some individual cases, but I've heard about it, that some people write there things that aren't completely true."} (F62)  
\end{quote}

Next to attributing possible errors in the Wikipedia articles to the lack of knowledge of ordinary editors, as shown in quotations above, some participants also noticed that articles can be deliberately manipulated. Moreover, older adults are aware that some topics, like biographies or politics, are more prone to this phenomenon.

\begin{quote}
\textit{"There may be some bias -- a biography may be written by someone's supporter or adversary."} (M87)
\end{quote}
\begin{quote}
\textit{"Credibility can be high, if there is no politics. I hope that's the case. I wouldn't dare even to question it."}(M75)
\end{quote}

\subsubsection{Editing policy}
Despite a strong opinion about possible problems with information credibility and malicious articles' manipulation, seniors had only limited knowledge about Wikipedia editing procedures. Participants did not know how Wikipedia works and who is responsible for editing articles. Most of them imagined that there is some sort of central council managing the whole system. 
\begin{quote}
\textit{"I do not know. Group of the people. Historians probably. Maybe tour guides from some city. Someone collects it together."} (F67)
\end{quote}

According to her some authority, competence and knowledge is required to edit the Wikipedia. The same conviction also appeared in the statements of others. Remain participants mentioned that they heard that \textit{anyone can be an editor} but none of them knew how to start. They also were convinced that still \textit{"some kind of body has to exist that reviews the submitted articles and decide whether to publish them or not"}. 
\begin{quote}
\textit{"I don't know to what extent this knowledge is verifiable by others."} (F62)
\end{quote}

\begin{quote}
\textit{"There is someone responsible for adding and filtering this knowledge."} (M75)
\end{quote}

\subsection{Editing Wikipedia by themselves}

\subsubsection{Searching for article}
First task was thought as a warm-up and was to make our participants familiar with the environment (i.a. room, computer, etc.). Majority of them put the name of the event in the Google's search bar and selected Wikipedia's web site from the obtained results. Without many problems all older adults managed to select the right article from Wikipedia disambiguation web page. Participants were familiar with the structure of articles and used both scrolling and clicking on internal links to navigate within the article. To sum up, seniors were attentive but, as expected, completed the first task without any problems. 

\subsubsection{Correcting factual information}
Second task required from participants switching from reading mode to editing mode. None of older adults taking part in the study figured out by themselves how to do it. Only with a prompt from research they were able to continue. 

Both wrong dates were visible on the first screen (without scrolling) -- one linked (so was highlighted in blue) in the first line, the article's summary above the table of contents, and the other to the right of the main text column, in the infobox war table. At first, the users noticed the incorrect date at the top of the article. The error was apparently easy to correct. Problematic, however, was the need for a "double edition" of the link -- link caption and target the destination to which it refers.

Vast majority of older adults was unaware of the fact that when they change the date from 1922 to 1920 referred only to the link, or rather to the destination to which it referred but not to the description of the link. So, they were surprised that their patch did not appear in the article when they approved with the \textit{Done} button. 

The necessity of double edition caused a confusion and led to the attempts that made things even worse. One user chose \textit{Edit mode} and clicked on the 1922 link to change the link. Next, in the pop-up window, instead of re-selecting Edit, she clicked on the 1922 link and opened new article dedicated to the year 1922 in a new tab (normally during browsing the wikipedia pages it opens in the same tab). The problem was that she did not notice that she was in another tab and tried, unsuccessfully, to return with the "back" button in the browser. The oldest of our respondents (M87) was the only one who quickly realized that the date and the link must be edited separately.
 
\begin{figure}
\centering
\includegraphics[height=220px]{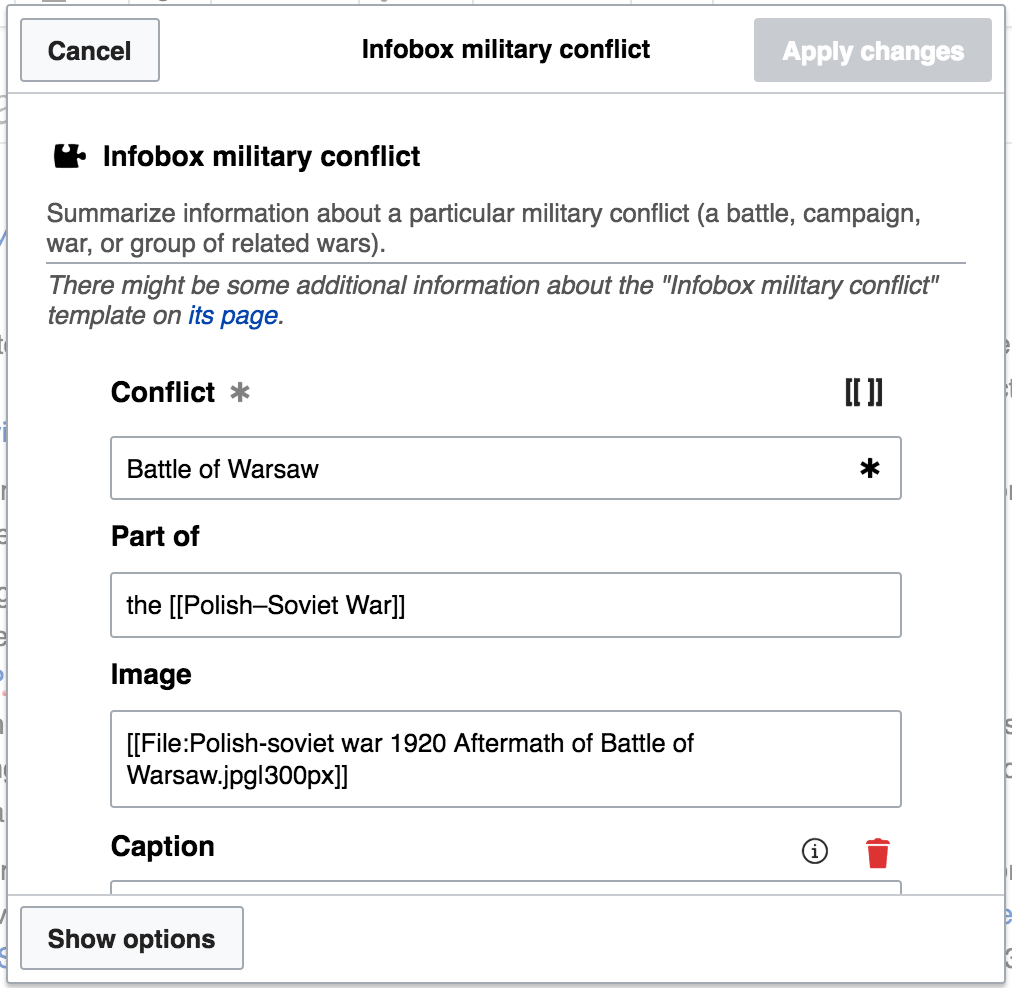}
\caption{Window for editing infoboxes.}
\label{fig-infobox}
\end{figure}

The attempt to correct the second error located in the info box reveals yet other problems with Wikipedia editing. It requires familiarizing with completely new editor that mixes the WYSWIG and code mode. First click on the noticed error in the infobox opens the window presented in Fig. \ref{fig-infobox} precluded by a notification displayed in bottom right corner of the web page: "Template. Generated from: infobox military". Wording of the pop-up was difficult to understand for participants. Especial two words confused older adults: "template" and "infobox". The difficulty in understanding the message was amplified by the fact that participants saw many new objects at the same time.

Older adults had also problems with understanding how the infobox editing window depicted in Fig. \ref{fig-infobox} should be mapped on the table they clicked on. Moreover, for most of them it was not clear that the infobox editing window can be scrolled. The biggest problem were, however, the brackets. None of participants guessed correctly what are they for. One of our respondents used curly brackets instead of square brackets because they are difficult to distinguish for people with defect of vision and are placed under the same key on the keyboard (at least for the US International keyboard layout). Yet another observed problem was that action buttons (i.e. "Apply changes" and "Cancel") are visually difficult to distinguish from the editing window. One of participants moved cursor to the top left corner when asked by researcher to save changes. This is an example of one of many habits acquired from using word processor (i.a. MS Word) observed during this study. Unfortunately, most of these habits did not lead to expected results and deepened feelings of unfamiliarity during editing the Wikipedia. 

\begin{figure}
\centering
\includegraphics[height=130px]{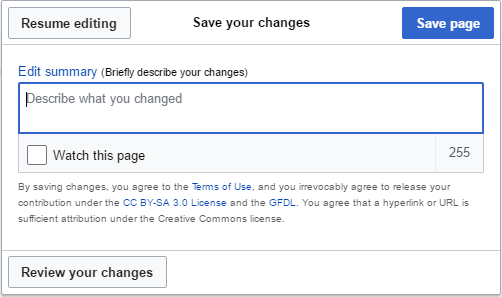}
\caption{Window which appears after click on the Save button. Editors are asked to describe changes they just made.}
\label{fig-changes}
\end{figure}

Unexpectedly for the participants of the study, clicking on the button \textit{Save} did not finish the job yet. Wikipedia editors are expected to describe each edition. The "Save your changes" window shown on Fig. \ref{fig-changes} was incomprehensible for participants. They did not understand the importance of describing changes (no justification for that was delivered next to the window) and even if they still wanted to do it neither hint nor examples were available to show how to make it correctly. Some minor issues like too distant placing of characters' counter made things even worse.

Even if seniors managed to overcome all obstacles with some advices from a researcher, they had problems with finding corrected information in the article after they switched back to reading mode. Only one participant noticed "Review your changes" option and was positively surprised by its usefulness. The majority noticed a centrally appearing message at the top of the page acknowledging that edition was saved, though a comment was also expressed: \textit{"I was not prepared for this. It should not disappear so quickly. I barely managed to notice it"} (F68). Despite many problems while correcting errors in the article seniors were satisfied with the results. 

\subsubsection{Adding internal link}

Seniors began to do the task with looking for the place in the article where they were expected to add a link. Next, they scrolled to the top of the page to click on the \textit{Edit} button. None of the participants noticed that they could also find the same button next to the title of the subsection. After switching to \textit{editing mode} participants were puzzled what they should do next. They needed a hint from a researcher to mark a piece of text. It turned out that selecting only few words without punctuation marks is a demanding task for older adults because require high psychomotor coordination (decreases with age) and a little practice (although seniors who participated in the study were quite experienced in comparison with people in similar age they were still not fluent in using touchpad and mouse). 

Moreover, participants did not know how typical link looks like (how many words it contains) though they used Wikipedia on regular basis. So they were unsure how many words they should mark. It turns out that reading articles does not make people aware of the templates used in Wikipedia articles and it is probably one of substantial barriers to overcome when people go from reader to contributor.

\begin{figure}
\centering
\includegraphics[height=130px]{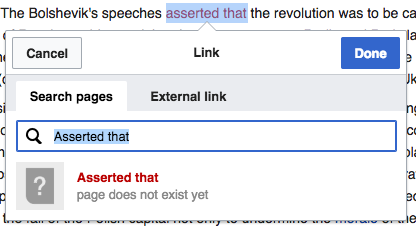}
\caption{Dialog window for adding link. By default the window open on the tab for adding internal links.}
\label{fig-link}
\end{figure}

The participants who did not notice the editors toolbar at the top of the screen, did not know what to do next. The toolbar is not sufficiently distinguishable from the color of the edited article. Contrast is low, and the whole bar is kept in shades of grey and it is difficult.

Even if they found the toolbar, they were puzzled what to choose. Some of them firstly looked at the \textit{Insert} or \textit{Cite} buttons. Finally, with some help from a researcher, the participants managed to click on the right button and opened the dialog window shown in the Fig. \ref{fig-link}. Probably because of the experience from the previous task they were not surprised by the fact that something is opening at the top of the web page but the window itself caused a lot of confusion. Majority of participants did not notice a search bar and started to scroll down the list of search results instead. Moreover, they were completely lost what this list of sentences was about and what they should do with it. Interestingly, seniors were strongly inclined to skip the first result accompanied with a picture and focused on remaining results. Some inconsistency in the Wikipedia editor also appeared when they clicked on one of the selected links. Link is created without any additional prompt and the dialog window disappears. Users expected that the editor should ask about confirmation of their choice or/and the Done \textit{button} had to be pressed.

Apparently they felt uneasy here, admitting that the next step they performed "on their own", and the reaction of the system happened to sigh with \textit{"O God!"}.

The lack of linguistic coherence in the names of "call-to-action" buttons was reported as a problem by participants. As can be seen in Fig. \ref{fig-infobox}, Fig. \ref{fig-changes}, and Fig. \ref{fig-link} action buttons are named differently: "Done", Save page", and "Apply changes". One of participants (F62) stated: \textit{"It was \textbf{Ready} there and here is \textbf{Apply}. Of course, it should be always the same."}.

\subsubsection{Creating new article}

Although the description of the task contained information that the article "literacy cafes" does not exist, a moderator suggested the participants that they have to start with verifying this information. After using the search bar in Wikipedia, seniors received a message from the Wikipedia web site that \textit{"Article 'Literary Cafe' does not exist in Wikipedia. You can create it (you have guides to help you) or recommend to create it by community."}. All participants were puzzled what to do next.

Messages about the lack of an article in Wikipedia are preceded by a pencil icon. The icon is disproportionately large in relation to the remaining content and icons used on the site, but unfortunately is not clickable, which seems to be a design flaw on the Wikipedia page\footnote{The English version of the Wikipedia looks slightly different and there is no such icon.}. Moreover, the title of the article that does not exist in the Wikipedia is highlighted in red and works as a link, and clicking on it launches the editor. At this point it is worth mentioning that most of the links are worded in a way that does not follow "call-to-action" guidelines. Instead of "click to editing" or "create" links were labeled with meaningless and passive words. 

After users finally clicked on the link that opens a web page for creating a new article, a welcome dialog box appeared with the invitation to edition, but there were no any further hints what to do except two flickering blue dots in the upper editing toolbar. Clicking on one of them opened a very friendly dialog box with a visual hint to include links to other wikipedia entries. The second one pointed to the advice that footnotes are important for acknowledging sources of information. These messages where helpful but more basic information where is the central window and how to start editing was missing.

On the right side two additional messages appeared. The first one was about anonymous edition of the Wikipedia, the second focused on guidelines for adding a new article to Wikipedia. These messages were not very helpful and mostly focused on prohibitions, furthermore none of participants noticed it.

Finally, older adults encountered some troubles during copying text from MS Word file, in which the content for the new article was prepared, to the Wikipedia web page, but it can be mostly attributed to limited proficiency in computer use.

\subsubsection{Creating own account}

Not all participants were interested in editing Wikipedia in the future. Some of them even did not want to create an account, which was required for completing homework. In total, six of ten participants returned to the Wikipedia editor at home. Others have given a variety of excuses to not to edit Wikipedia - i.a. inadequate knowledge, reluctant to write, age, fear of responsibility, loss of privacy, and finally, laziness.  

\begin{quote}
\textit{"For me it's strange that there is written: you don't need any special skills. It's very strange to me. I think that knowledge is necessary. And special one if somenone writes things concerning some subject. I was never good in writing papers for Polish classes."} (F68)

\textit{"For me it is already rather impossible. Let's leave it to the young ones."} (F67)

\textit{"I'm not into editing, I'm too less qualified. I don't have either skills nor knowledge. (...) There are two important things here -- first of all, technical aspect of the matter, and second of all, what I want to introduce - knowledge itself. It has to be a predestined person withi this knowledge. It can be two people - one with knowledge and secon with the technique of introducing. It's difficult for me too, to use it. I can forget everything tomorrow - that's the charm of an elder person.".} (M75)
\end{quote}

Those from the participants who decided to try to edit Wikipedia in the future had to create an account. If they managed to find option "create account" in the top menu they started immediately filling up form. If someone started from the left side menu, things were a bit more complicated and some assist from a researcher was required. Some of older adults were 
reluctant to give personal information during the registration process -- \textit{"I don't like it in the Internet at all. Creating account always means giving private data."} (F68) -- but other did not have any problem with that -- \textit{"An email address is optional. You can give it but you don't have to. I have nothing to hide. All what I do is newer done under cover."} (M68)

The only serious difficulty during filling up the registration form was CAPTCHA. In general users do not like it -- \textit{"It's the worst what can be. I don't like it."} (F62) -- and not always understand what they should do -- \textit{"What with this code? What they call code? Is it password?"} (M68). Frustration was even further amplified by sights problem that are quite often among seniors -- \textit{"I think I'm a robot, because I cannot see it. Is this 'l'? How do I know it?"} (M75). Many of participants admitted that if they were to go through it alone, they would have \textit{"threw it down long ago"} and abandon creating an account.

However, some participants were really curious about the possibilities of Wikipedia and were motivated to try, despite of encountered difficulties. 

\begin{quote}
\textit{ "It is a new world which people do not explore by themselves because of laziness. (...) Meeting with the researcher provoke me to go deeper into it and enjoy it."} (M68)

\textit{"I am delighted. I did not know that such things could be done."} (F64)
\end{quote}

\section{Discussion}

The quotations extracted from the experiment show that the idea of editing Wikipedia is appealing to, at least, some of older adults taking part in the study. Nevertheless, even for the elderly with above average computer skills and high intrinsic motivation present version of the Wikipedia editor is strongly discouraging. The most important obstacles are listed below accompanied by recommendation and suggestions how they can be addressed.

The very first issue appeared when the elderly sat in front of the Wikipedia editor (and web site in general) was low contrast and limited selection of colors -- mostly shades of gray. It was particularly evident when participants opened the editor window for the first time. Most of them did not notice the top toolbar. Neither the toolbar  nor the buttons are differentiated enough from the rest of the web page (black text on white background are also used in the wikipedia articles). Even if color change is not possible due to the habits and resistance of the Wikipedians community, a special version of the web page could be prepared for people requiring higher contrast.

The Wiki platform was developed incrementally. New functionalities were (and are) added gradually with the emerging needs. Therefore, in many places the user interface is not coherent. The same or very similar icons have different meaning. Sometimes the dialog window appears above and sometimes below the item being edited. Moreover, there were also cases where the dialog window jumped out outside the part of the web page currently rendered in the web browser window. Coherence was also lacking at the linguistic level. This was most apparent in the buttons' labels. 

Often even when the selection of words seemed appropriate and particular words were consequently use across the whole Wikipedia, these words were hard to understand even by the well educated seniors participating in this study. The most striking example is the word \textit{"multimedia"} (pol. \textit{"multimedia"}). Older adults participating in the study intentionally skipped this option when asked to add picture to the article. For them word \textit{multimedia} was synonymous with \textit{video} but not with picture.

Although the research concerned the Polish version of Wikipedia, a quick check reveals that the same problem is present, to some extend, also in other language versions. On one hand changes in the used words are the easiest to implement as it does not require any modification in the code. On the other hand it cannot be done universally for the whole Wikipedia and requires "language by language" approach.

Positive feedback (information from the system that action was successful) is one of the most important requirements for designing a user-friendly system. Wikipedia tries to follow it but on some occasions it fails. For most actions there are message windows appearing on the top of the page for few seconds and disappear. Most of the participants had problem with completing the whole cycle during that time period: noticing, reading, and understanding. It lefts users with the feeling that they missed something and further undermined their self-confidence. A design pattern requiring an acknowledgement of all notification would be probably better suited to the needs of the elderly but could be, at the same time, cumbersome for hardcore editors.

Even those participants who were quite successful in completing tasks without (or with limited) assist from researcher made unintentional, and usually also unnoticed by them, changes in the editing articles. Two most common errors were: adding new empty paragraph by clicking above existing one and adding random links. Of course, all these unintentional edits might be removed by another Wikipedia users but it imposes additional work on already overworked wikipedia editors. Therefore an implementation of additional mechanism for limiting such error is necessary. It can be done either by asking editors (maybe only some unexperienced editors) to once more acknowledge each change or by limiting options to only a subset needed for completing a prior defined task (e.g. if user intends to add link all options related to editing structure of the article, adding multimedia or footnotes might be hidden by default).

In general, users expect that all programs and web sites will behave in the same way. A convergence between web sites, desktop application and mobile application in terms of user interface can be currently observed. The best example of this process is Microsoft Word Online which deliver almost the same feeling like the desktop application. The development of the Wikipedia editor has mostly been done in times when the same level of adequacy was hard to achieve because of technical limitation, but nowadays it is completely possible. The most striking example is a context menu displayed after clicking on the left mouse button. Participants expected to see a context menu related to the Wikipedia editing process but seen a web browser context menu instead. It is crucial because for older adults familiarity has been shown to be one of the most important factor that affects acceptance. 

During the study seniors were eager to ask questions and look for guidance from the researcher. Moreover, opinions expressed during the editing indicate that opportunity to consult own doubts was key to maintaining motivation. An live chat is a possible solution to assure some assist for beginner editors but
it requires a lot of effort from an experienced editors to provide support. Recent advances and successful deployments of the chat bots might be an way to limit necessary work of people. 

\section{Conclusion}

Despite many encountered problems, older adults were positively oriented to editing Wikipedia. Of course only a fraction of the participants returned to the Wikipedia editor by themselves at home (three out of ten) but contributing to the Wikipedia is and in foreseeable future will be an elitist activity.  

It needs to be highlighted that although the recommendation it the previous section focus on changes that should be done in Wikipedia, many of the issues raised by the participants might albo be solved either by extensive training or more regular exercises.

As a part of further research, a preparation of the online learning course about contributing to the Wikipedia directed to older adults is planned. The on-line course enriched with some off-line meetings will be used to build a community of editors. Important research question is also how contributing to the Wikipedia influence a well-being of older adults in the short and long term perspective.  

\section{Acknowledgments}
This project has received funding from the European Union's Horizon 2020 research and innovation programme under the Marie Sklodowska-Curie grant agreement No 690962

\bibliographystyle{ACM-Reference-Format}
\bibliography{sample} 


\begin{thebibliography}{00}


\ifx \showCODEN    \undefined \def \showCODEN     #1{\unskip}     \fi
\ifx \showDOI      \undefined \def \showDOI       #1{#1}\fi
\ifx \showISBNx    \undefined \def \showISBNx     #1{\unskip}     \fi
\ifx \showISBNxiii \undefined \def \showISBNxiii  #1{\unskip}     \fi
\ifx \showISSN     \undefined \def \showISSN      #1{\unskip}     \fi
\ifx \showLCCN     \undefined \def \showLCCN      #1{\unskip}     \fi
\ifx \shownote     \undefined \def \shownote      #1{#1}          \fi
\ifx \showarticletitle \undefined \def \showarticletitle #1{#1}   \fi
\ifx \showURL      \undefined \def \showURL       {\relax}        \fi
\providecommand\bibfield[2]{#2}
\providecommand\bibinfo[2]{#2}
\providecommand\natexlab[1]{#1}
\providecommand\showeprint[2][]{arXiv:#2}

\bibitem[\protect\citeauthoryear{Antin}{Antin}{2011}]%
        {antin2011my}
\bibfield{author}{\bibinfo{person}{Judd Antin}.}
  \bibinfo{year}{2011}\natexlab{}.
\newblock \showarticletitle{My kind of people?: perceptions about wikipedia
  contributors and their motivations}. In \bibinfo{booktitle}{{\em Proceedings
  of the SIGCHI Conference on Human Factors in Computing Systems}}. ACM,
  \bibinfo{pages}{3411--3420}.
\newblock


\bibitem[\protect\citeauthoryear{Antin, Yee, Cheshire, and Nov}{Antin
  et~al\mbox{.}}{2011}]%
        {antin2011gender}
\bibfield{author}{\bibinfo{person}{Judd Antin}, \bibinfo{person}{Raymond Yee},
  \bibinfo{person}{Coye Cheshire}, {and} \bibinfo{person}{Oded Nov}.}
  \bibinfo{year}{2011}\natexlab{}.
\newblock \showarticletitle{Gender differences in Wikipedia editing}. In
  \bibinfo{booktitle}{{\em Proceedings of the 7th International Symposium on
  Wikis and Open Collaboration}}. ACM, \bibinfo{pages}{11--14}.
\newblock


\bibitem[\protect\citeauthoryear{Brake}{Brake}{2014}]%
        {brake2014we}
\bibfield{author}{\bibinfo{person}{David~R Brake}.}
  \bibinfo{year}{2014}\natexlab{}.
\newblock \showarticletitle{Are we all online content creators now? Web 2.0 and
  digital divides}.
\newblock \bibinfo{journal}{{\em Journal of Computer-Mediated Communication\/}}
  \bibinfo{volume}{19}, \bibinfo{number}{3} (\bibinfo{year}{2014}),
  \bibinfo{pages}{591--609}.
\newblock


\bibitem[\protect\citeauthoryear{Brewer, Morris, and Piper}{Brewer
  et~al\mbox{.}}{2016}]%
        {brewer2016would}
\bibfield{author}{\bibinfo{person}{Robin Brewer},
  \bibinfo{person}{Meredith~Ringel Morris}, {and} \bibinfo{person}{Anne~Marie
  Piper}.} \bibinfo{year}{2016}\natexlab{}.
\newblock \showarticletitle{Why would anybody do this?: Understanding Older
  Adults' Motivations and Challenges in Crowd Work}. In
  \bibinfo{booktitle}{{\em Proceedings of the 2016 CHI Conference on Human
  Factors in Computing Systems}}. ACM, \bibinfo{pages}{2246--2257}.
\newblock


\bibitem[\protect\citeauthoryear{Brewer and Piper}{Brewer and Piper}{2016}]%
        {Brewer:2016:TLR:2858036.2858379}
\bibfield{author}{\bibinfo{person}{Robin Brewer} {and}
  \bibinfo{person}{Anne~Marie Piper}.} \bibinfo{year}{2016}\natexlab{}.
\newblock \showarticletitle{"Tell It Like It Really Is": A Case of Online
  Content Creation and Sharing Among Older Adult Bloggers}. In
  \bibinfo{booktitle}{{\em Proceedings of the 2016 CHI Conference on Human
  Factors in Computing Systems}} {\em (\bibinfo{series}{CHI '16})}.
  \bibinfo{publisher}{ACM}, \bibinfo{address}{New York, NY, USA},
  \bibinfo{pages}{5529--5542}.
\newblock
\showISBNx{978-1-4503-3362-7}
\showDOI{%
\url{https://doi.org/10.1145/2858036.2858379}}


\bibitem[\protect\citeauthoryear{Buzzi, Buzzi, Leporini, and Senette}{Buzzi
  et~al\mbox{.}}{2008}]%
        {buzzi2008making}
\bibfield{author}{\bibinfo{person}{M~Claudia Buzzi}, \bibinfo{person}{Marina
  Buzzi}, \bibinfo{person}{Barbara Leporini}, {and} \bibinfo{person}{Caterina
  Senette}.} \bibinfo{year}{2008}\natexlab{}.
\newblock \showarticletitle{Making Wikipedia editing easier for the blind}. In
  \bibinfo{booktitle}{{\em Proceedings of the 5th Nordic conference on
  Human-computer interaction: building bridges}}. ACM,
  \bibinfo{pages}{423--426}.
\newblock


\bibitem[\protect\citeauthoryear{Chang, Kumar, Gilbert, and Terveen}{Chang
  et~al\mbox{.}}{2014}]%
        {chang2014specialization}
\bibfield{author}{\bibinfo{person}{Shuo Chang}, \bibinfo{person}{Vikas Kumar},
  \bibinfo{person}{Eric Gilbert}, {and} \bibinfo{person}{Loren~G Terveen}.}
  \bibinfo{year}{2014}\natexlab{}.
\newblock \showarticletitle{Specialization, homophily, and gender in a social
  curation site: findings from pinterest}. In \bibinfo{booktitle}{{\em
  Proceedings of the 17th ACM conference on Computer supported cooperative work
  \& social computing}}. ACM, \bibinfo{pages}{674--686}.
\newblock


\bibitem[\protect\citeauthoryear{Ciampaglia and Taraborelli}{Ciampaglia and
  Taraborelli}{2015}]%
        {ciampaglia2015moodbar}
\bibfield{author}{\bibinfo{person}{Giovanni~Luca Ciampaglia} {and}
  \bibinfo{person}{Dario Taraborelli}.} \bibinfo{year}{2015}\natexlab{}.
\newblock \showarticletitle{MoodBar: Increasing new user retention in Wikipedia
  through lightweight socialization}. In \bibinfo{booktitle}{{\em Proceedings
  of the 18th ACM Conference on Computer Supported Cooperative Work \& Social
  Computing}}. ACM, \bibinfo{pages}{734--742}.
\newblock


\bibitem[\protect\citeauthoryear{Dickinson, Arnott, and Prior}{Dickinson
  et~al\mbox{.}}{2007}]%
        {dickinson2007methods}
\bibfield{author}{\bibinfo{person}{Anna Dickinson}, \bibinfo{person}{John
  Arnott}, {and} \bibinfo{person}{Suzanne Prior}.}
  \bibinfo{year}{2007}\natexlab{}.
\newblock \showarticletitle{Methods for human--computer interaction research
  with older people}.
\newblock \bibinfo{journal}{{\em Behaviour \& Information Technology\/}}
  \bibinfo{volume}{26}, \bibinfo{number}{4} (\bibinfo{year}{2007}),
  \bibinfo{pages}{343--352}.
\newblock


\bibitem[\protect\citeauthoryear{Dickinson and Gregor}{Dickinson and
  Gregor}{2006}]%
        {dickinson2006computer}
\bibfield{author}{\bibinfo{person}{Anna Dickinson} {and} \bibinfo{person}{Peter
  Gregor}.} \bibinfo{year}{2006}\natexlab{}.
\newblock \showarticletitle{Computer use has no demonstrated impact on the
  well-being of older adults}.
\newblock \bibinfo{journal}{{\em International Journal of Human-Computer
  Studies\/}} \bibinfo{volume}{64}, \bibinfo{number}{8} (\bibinfo{year}{2006}),
  \bibinfo{pages}{744--753}.
\newblock


\bibitem[\protect\citeauthoryear{Farzan and Kraut}{Farzan and Kraut}{2013}]%
        {Farzan:2013:WCE:2470654.2470765}
\bibfield{author}{\bibinfo{person}{Rosta Farzan} {and}
  \bibinfo{person}{Robert~E. Kraut}.} \bibinfo{year}{2013}\natexlab{}.
\newblock \showarticletitle{Wikipedia Classroom Experiment: Bidirectional
  Benefits of Students' Engagement in Online Production Communities}. In
  \bibinfo{booktitle}{{\em Proceedings of the SIGCHI Conference on Human
  Factors in Computing Systems}} {\em (\bibinfo{series}{CHI '13})}.
  \bibinfo{publisher}{ACM}, \bibinfo{address}{New York, NY, USA},
  \bibinfo{pages}{783--792}.
\newblock
\showISBNx{978-1-4503-1899-0}
\showDOI{%
\url{https://doi.org/10.1145/2470654.2470765}}


\bibitem[\protect\citeauthoryear{Forte, Antin, Bardzell, Honeywell, Riedl, and
  Stierch}{Forte et~al\mbox{.}}{2012}]%
        {forte2012some}
\bibfield{author}{\bibinfo{person}{Andrea Forte}, \bibinfo{person}{Judd Antin},
  \bibinfo{person}{Shaowen Bardzell}, \bibinfo{person}{Leigh Honeywell},
  \bibinfo{person}{John Riedl}, {and} \bibinfo{person}{Sarah Stierch}.}
  \bibinfo{year}{2012}\natexlab{}.
\newblock \showarticletitle{Some of all human knowledge: gender and
  participation in peer production}. In \bibinfo{booktitle}{{\em Proceedings of
  the ACM 2012 conference on Computer Supported Cooperative Work Companion}}.
  ACM, \bibinfo{pages}{33--36}.
\newblock


\bibitem[\protect\citeauthoryear{Greenfield and Marks}{Greenfield and
  Marks}{2004}]%
        {greenfield2004formal}
\bibfield{author}{\bibinfo{person}{Emily~A Greenfield} {and}
  \bibinfo{person}{Nadine~F Marks}.} \bibinfo{year}{2004}\natexlab{}.
\newblock \showarticletitle{Formal volunteering as a protective factor for
  older adults' psychological well-being}.
\newblock \bibinfo{journal}{{\em The Journals of Gerontology Series B:
  Psychological Sciences and Social Sciences\/}} \bibinfo{volume}{59},
  \bibinfo{number}{5} (\bibinfo{year}{2004}), \bibinfo{pages}{S258--S264}.
\newblock


\bibitem[\protect\citeauthoryear{Halfaker, Keyes, and Taraborelli}{Halfaker
  et~al\mbox{.}}{2013}]%
        {halfaker2013making}
\bibfield{author}{\bibinfo{person}{Aaron Halfaker}, \bibinfo{person}{Oliver
  Keyes}, {and} \bibinfo{person}{Dario Taraborelli}.}
  \bibinfo{year}{2013}\natexlab{}.
\newblock \showarticletitle{Making peripheral participation legitimate: reader
  engagement experiments in wikipedia}. In \bibinfo{booktitle}{{\em Proceedings
  of the 2013 conference on Computer supported cooperative work}}. ACM,
  \bibinfo{pages}{849--860}.
\newblock


\bibitem[\protect\citeauthoryear{Hao}{Hao}{2008}]%
        {hao2008productive}
\bibfield{author}{\bibinfo{person}{Yanni Hao}.}
  \bibinfo{year}{2008}\natexlab{}.
\newblock \showarticletitle{Productive activities and psychological well-being
  among older adults}.
\newblock \bibinfo{journal}{{\em The Journals of Gerontology Series B:
  Psychological Sciences and Social Sciences\/}} \bibinfo{volume}{63},
  \bibinfo{number}{2} (\bibinfo{year}{2008}), \bibinfo{pages}{S64--S72}.
\newblock


\bibitem[\protect\citeauthoryear{Hiyama, Nagai, Hirose, Kobayashi, and
  Takagi}{Hiyama et~al\mbox{.}}{2013}]%
        {hiyama2013question}
\bibfield{author}{\bibinfo{person}{Atsushi Hiyama}, \bibinfo{person}{Yuki
  Nagai}, \bibinfo{person}{Michitaka Hirose}, \bibinfo{person}{Masatomo
  Kobayashi}, {and} \bibinfo{person}{Hironobu Takagi}.}
  \bibinfo{year}{2013}\natexlab{}.
\newblock \showarticletitle{Question first: Passive interaction model for
  gathering experience and knowledge from the elderly}. In
  \bibinfo{booktitle}{{\em Pervasive Computing and Communications Workshops
  (PERCOM Workshops), 2013 IEEE International Conference on}}. IEEE,
  \bibinfo{pages}{151--156}.
\newblock


\bibitem[\protect\citeauthoryear{Ibarra, Korovina, Baez, Barysheva, Marchese,
  Cernuzzi, and Casati}{Ibarra et~al\mbox{.}}{}]%
        {ibarratools}
\bibfield{author}{\bibinfo{person}{Francisco Ibarra}, \bibinfo{person}{Olga
  Korovina}, \bibinfo{person}{Marcos Baez}, \bibinfo{person}{Galina Barysheva},
  \bibinfo{person}{Maurizio Marchese}, \bibinfo{person}{Luca Cernuzzi}, {and}
  \bibinfo{person}{Fabio Casati}.}
\newblock \showarticletitle{Tools Enabling Online Contributions by Older
  Adults}.
\newblock  (\bibinfo{year}{????}).
\newblock


\bibitem[\protect\citeauthoryear{Itoko, Arita, Kobayashi, and Takagi}{Itoko
  et~al\mbox{.}}{2014}]%
        {itoko2014involving}
\bibfield{author}{\bibinfo{person}{Toshinari Itoko}, \bibinfo{person}{Shoma
  Arita}, \bibinfo{person}{Masatomo Kobayashi}, {and} \bibinfo{person}{Hironobu
  Takagi}.} \bibinfo{year}{2014}\natexlab{}.
\newblock \showarticletitle{Involving senior workers in crowdsourced
  proofreading}. In \bibinfo{booktitle}{{\em International Conference on
  Universal Access in Human-Computer Interaction}}. Springer,
  \bibinfo{pages}{106--117}.
\newblock


\bibitem[\protect\citeauthoryear{Jankowski-Lorek, Nielek, Wierzbicki, and
  Zieli{\'n}ski}{Jankowski-Lorek et~al\mbox{.}}{2014}]%
        {jankowski2014predicting}
\bibfield{author}{\bibinfo{person}{Micha{\l} Jankowski-Lorek},
  \bibinfo{person}{Rados{\l}aw Nielek}, \bibinfo{person}{Adam Wierzbicki},
  {and} \bibinfo{person}{Kazimierz Zieli{\'n}ski}.}
  \bibinfo{year}{2014}\natexlab{}.
\newblock \showarticletitle{Predicting controversy of wikipedia articles using
  the article feedback tool}. In \bibinfo{booktitle}{{\em Proceedings of the
  2014 International Conference on Social Computing}}. ACM,
  \bibinfo{pages}{22}.
\newblock


\bibitem[\protect\citeauthoryear{Kobayashi, Arita, Itoko, Saito, and
  Takagi}{Kobayashi et~al\mbox{.}}{2015}]%
        {kobayashi2015motivating}
\bibfield{author}{\bibinfo{person}{Masatomo Kobayashi}, \bibinfo{person}{Shoma
  Arita}, \bibinfo{person}{Toshinari Itoko}, \bibinfo{person}{Shin Saito},
  {and} \bibinfo{person}{Hironobu Takagi}.} \bibinfo{year}{2015}\natexlab{}.
\newblock \showarticletitle{Motivating multi-generational crowd workers in
  social-purpose work}. In \bibinfo{booktitle}{{\em Proceedings of the 18th ACM
  Conference on Computer Supported Cooperative Work \& Social Computing}}. ACM,
  \bibinfo{pages}{1813--1824}.
\newblock


\bibitem[\protect\citeauthoryear{Kobayashi, Ishihara, Itoko, Takagi, and
  Asakawa}{Kobayashi et~al\mbox{.}}{2013}]%
        {kobayashi2013age}
\bibfield{author}{\bibinfo{person}{Masatomo Kobayashi},
  \bibinfo{person}{Tatsuya Ishihara}, \bibinfo{person}{Toshinari Itoko},
  \bibinfo{person}{Hironobu Takagi}, {and} \bibinfo{person}{Chieko Asakawa}.}
  \bibinfo{year}{2013}\natexlab{}.
\newblock \showarticletitle{Age-based task specialization for crowdsourced
  proofreading}. In \bibinfo{booktitle}{{\em International Conference on
  Universal Access in Human-Computer Interaction}}. Springer,
  \bibinfo{pages}{104--112}.
\newblock


\bibitem[\protect\citeauthoryear{Kuznetsov}{Kuznetsov}{2006}]%
        {Kuznetsov:2006:MCW:1215942.1215943}
\bibfield{author}{\bibinfo{person}{Stacey Kuznetsov}.}
  \bibinfo{year}{2006}\natexlab{}.
\newblock \showarticletitle{Motivations of Contributors to Wikipedia}.
\newblock \bibinfo{journal}{{\em SIGCAS Comput. Soc.\/}} \bibinfo{volume}{36},
  \bibinfo{number}{2}, Article \bibinfo{articleno}{1} (\bibinfo{date}{June}
  \bibinfo{year}{2006}).
\newblock
\showISSN{0095-2737}
\showDOI{%
\url{https://doi.org/10.1145/1215942.1215943}}


\bibitem[\protect\citeauthoryear{Lam, Uduwage, Dong, Sen, Musicant, Terveen,
  and Riedl}{Lam et~al\mbox{.}}{2011}]%
        {lam2011wp}
\bibfield{author}{\bibinfo{person}{Shyong Tony~K Lam},
  \bibinfo{person}{Anuradha Uduwage}, \bibinfo{person}{Zhenhua Dong},
  \bibinfo{person}{Shilad Sen}, \bibinfo{person}{David~R Musicant},
  \bibinfo{person}{Loren Terveen}, {and} \bibinfo{person}{John Riedl}.}
  \bibinfo{year}{2011}\natexlab{}.
\newblock \showarticletitle{WP: clubhouse?: an exploration of Wikipedia's
  gender imbalance}. In \bibinfo{booktitle}{{\em Proceedings of the 7th
  international symposium on Wikis and open collaboration}}. ACM,
  \bibinfo{pages}{1--10}.
\newblock


\bibitem[\protect\citeauthoryear{Lee and Seo}{Lee and Seo}{2016}]%
        {lee2016crowdsourcing}
\bibfield{author}{\bibinfo{person}{Jung Lee} {and} \bibinfo{person}{DongBack
  Seo}.} \bibinfo{year}{2016}\natexlab{}.
\newblock \showarticletitle{Crowdsourcing not all sourced by the crowd: An
  observation on the behavior of Wikipedia participants}.
\newblock \bibinfo{journal}{{\em Technovation\/}} (\bibinfo{year}{2016}).
\newblock


\bibitem[\protect\citeauthoryear{Lum and Lightfoot}{Lum and Lightfoot}{2005}]%
        {lum2005effects}
\bibfield{author}{\bibinfo{person}{Terry~Y Lum} {and}
  \bibinfo{person}{Elizabeth Lightfoot}.} \bibinfo{year}{2005}\natexlab{}.
\newblock \showarticletitle{The effects of volunteering on the physical and
  mental health of older people}.
\newblock \bibinfo{journal}{{\em Research on aging\/}} \bibinfo{volume}{27},
  \bibinfo{number}{1} (\bibinfo{year}{2005}), \bibinfo{pages}{31--55}.
\newblock


\bibitem[\protect\citeauthoryear{Morgan, Bouterse, Walls, and Stierch}{Morgan
  et~al\mbox{.}}{2013}]%
        {morgan2013tea}
\bibfield{author}{\bibinfo{person}{Jonathan~T Morgan}, \bibinfo{person}{Siko
  Bouterse}, \bibinfo{person}{Heather Walls}, {and} \bibinfo{person}{Sarah
  Stierch}.} \bibinfo{year}{2013}\natexlab{}.
\newblock \showarticletitle{Tea and sympathy: crafting positive new user
  experiences on wikipedia}. In \bibinfo{booktitle}{{\em Proceedings of the
  2013 conference on Computer supported cooperative work}}. ACM,
  \bibinfo{pages}{839--848}.
\newblock


\bibitem[\protect\citeauthoryear{Morrow-Howell}{Morrow-Howell}{2010}]%
        {morrow2010volunteering}
\bibfield{author}{\bibinfo{person}{Nancy Morrow-Howell}.}
  \bibinfo{year}{2010}\natexlab{}.
\newblock \showarticletitle{Volunteering in later life: Research frontiers}.
\newblock \bibinfo{journal}{{\em The Journals of Gerontology Series B:
  Psychological Sciences and Social Sciences\/}} \bibinfo{volume}{65},
  \bibinfo{number}{4} (\bibinfo{year}{2010}), \bibinfo{pages}{461--469}.
\newblock


\bibitem[\protect\citeauthoryear{Morrow-Howell, Hinterlong, Rozario, and
  Tang}{Morrow-Howell et~al\mbox{.}}{2003}]%
        {morrow2003effects}
\bibfield{author}{\bibinfo{person}{Nancy Morrow-Howell}, \bibinfo{person}{Jim
  Hinterlong}, \bibinfo{person}{Philip~A Rozario}, {and}
  \bibinfo{person}{Fengyan Tang}.} \bibinfo{year}{2003}\natexlab{}.
\newblock \showarticletitle{Effects of volunteering on the well-being of older
  adults}.
\newblock \bibinfo{journal}{{\em The Journals of Gerontology Series B:
  Psychological Sciences and Social Sciences\/}} \bibinfo{volume}{58},
  \bibinfo{number}{3} (\bibinfo{year}{2003}), \bibinfo{pages}{S137--S145}.
\newblock


\bibitem[\protect\citeauthoryear{Narayan, Orlowitz, Morgan, and Shaw}{Narayan
  et~al\mbox{.}}{2015}]%
        {narayan2015effects}
\bibfield{author}{\bibinfo{person}{Sneha Narayan}, \bibinfo{person}{Jake
  Orlowitz}, \bibinfo{person}{Jonathan~T Morgan}, {and} \bibinfo{person}{Aaron
  Shaw}.} \bibinfo{year}{2015}\natexlab{}.
\newblock \showarticletitle{Effects of a Wikipedia Orientation Game on New User
  Edits}. In \bibinfo{booktitle}{{\em Proceedings of the 18th ACM Conference
  Companion on Computer Supported Cooperative Work \& Social Computing}}. ACM,
  \bibinfo{pages}{263--266}.
\newblock


\bibitem[\protect\citeauthoryear{Pan, Miao, Yu, Leung, and Chin}{Pan
  et~al\mbox{.}}{2015}]%
        {pan2015effects}
\bibfield{author}{\bibinfo{person}{Zhengxiang Pan}, \bibinfo{person}{Chunyan
  Miao}, \bibinfo{person}{Han Yu}, \bibinfo{person}{Cyril Leung}, {and}
  \bibinfo{person}{Jing~Jih Chin}.} \bibinfo{year}{2015}\natexlab{}.
\newblock \showarticletitle{The effects of familiarity design on the adoption
  of wellness games by the elderly}. In \bibinfo{booktitle}{{\em 2015
  IEEE/WIC/ACM International Conference on Web Intelligence and Intelligent
  Agent Technology (WI-IAT)}}, Vol.~\bibinfo{volume}{2}. IEEE,
  \bibinfo{pages}{387--390}.
\newblock


\bibitem[\protect\citeauthoryear{Roth, Davis, and Carver}{Roth
  et~al\mbox{.}}{2013}]%
        {roth2013assigning}
\bibfield{author}{\bibinfo{person}{Amy Roth}, \bibinfo{person}{Rochelle Davis},
  {and} \bibinfo{person}{Brian Carver}.} \bibinfo{year}{2013}\natexlab{}.
\newblock \showarticletitle{Assigning Wikipedia editing: Triangulation toward
  understanding university student engagement}.
\newblock \bibinfo{journal}{{\em First Monday\/}} \bibinfo{volume}{18},
  \bibinfo{number}{6} (\bibinfo{year}{2013}).
\newblock


\bibitem[\protect\citeauthoryear{Seiderer, Hammer, Andre, Mayr, and
  Rist}{Seiderer et~al\mbox{.}}{2015}]%
        {Seiderer:2015:EDI:2750511.2750514}
\bibfield{author}{\bibinfo{person}{Andreas Seiderer}, \bibinfo{person}{Stephan
  Hammer}, \bibinfo{person}{Elisabeth Andre}, \bibinfo{person}{Marcus Mayr},
  {and} \bibinfo{person}{Thomas Rist}.} \bibinfo{year}{2015}\natexlab{}.
\newblock \showarticletitle{Exploring Digital Image Frames for Lifestyle
  Intervention to Improve Well-being of Older Adults}. In
  \bibinfo{booktitle}{{\em Proceedings of the 5th International Conference on
  Digital Health 2015}} {\em (\bibinfo{series}{DH '15})}.
  \bibinfo{publisher}{ACM}, \bibinfo{address}{New York, NY, USA},
  \bibinfo{pages}{71--78}.
\newblock
\showISBNx{978-1-4503-3492-1}
\showDOI{%
\url{https://doi.org/10.1145/2750511.2750514}}


\bibitem[\protect\citeauthoryear{Senette, Buzzi, Buzzi, and Leporini}{Senette
  et~al\mbox{.}}{2009}]%
        {senette2009enhancing}
\bibfield{author}{\bibinfo{person}{Caterina Senette},
  \bibinfo{person}{Maria~Claudia Buzzi}, \bibinfo{person}{Marina Buzzi}, {and}
  \bibinfo{person}{Barbara Leporini}.} \bibinfo{year}{2009}\natexlab{}.
\newblock \showarticletitle{Enhancing Wikipedia Editing with WAI-ARIA}. In
  \bibinfo{booktitle}{{\em Symposium of the Austrian HCI and Usability
  Engineering Group}}. Springer, \bibinfo{pages}{159--177}.
\newblock


\bibitem[\protect\citeauthoryear{Turek, Wierzbicki, Nielek, and Datta}{Turek
  et~al\mbox{.}}{2011}]%
        {turek2011wikiteams}
\bibfield{author}{\bibinfo{person}{Piotr Turek}, \bibinfo{person}{Adam
  Wierzbicki}, \bibinfo{person}{Rados{\l}aw Nielek}, {and}
  \bibinfo{person}{Anwitaman Datta}.} \bibinfo{year}{2011}\natexlab{}.
\newblock \showarticletitle{Wikiteams: How do they achieve success?}
\newblock \bibinfo{journal}{{\em IEEE Potentials\/}} \bibinfo{volume}{30},
  \bibinfo{number}{5} (\bibinfo{year}{2011}), \bibinfo{pages}{15--20}.
\newblock


\bibitem[\protect\citeauthoryear{Turek, Wierzbicki, Nielek, Hupa, and
  Datta}{Turek et~al\mbox{.}}{2010}]%
        {turek2010learning}
\bibfield{author}{\bibinfo{person}{Piotr Turek}, \bibinfo{person}{Adam
  Wierzbicki}, \bibinfo{person}{Radosaw Nielek}, \bibinfo{person}{Albert Hupa},
  {and} \bibinfo{person}{Anwitaman Datta}.} \bibinfo{year}{2010}\natexlab{}.
\newblock \showarticletitle{Learning about the quality of teamwork from
  wikiteams}. In \bibinfo{booktitle}{{\em Social Computing (SocialCom), 2010
  IEEE Second International Conference on}}. IEEE, \bibinfo{pages}{17--24}.
\newblock


\bibitem[\protect\citeauthoryear{Vora, Komura, and Team}{Vora
  et~al\mbox{.}}{2010}]%
        {vora2010n00b}
\bibfield{author}{\bibinfo{person}{Parul Vora}, \bibinfo{person}{Naoko Komura},
  {and} \bibinfo{person}{Stanton~Usability Team}.}
  \bibinfo{year}{2010}\natexlab{}.
\newblock \showarticletitle{The n00b Wikipedia editing experience}. In
  \bibinfo{booktitle}{{\em Proceedings of the 6th International Symposium on
  Wikis and Open Collaboration}}. ACM, \bibinfo{pages}{36}.
\newblock


\bibitem[\protect\citeauthoryear{Wierzbicki, Turek, and Nielek}{Wierzbicki
  et~al\mbox{.}}{2010}]%
        {wierzbicki2010learning}
\bibfield{author}{\bibinfo{person}{Adam Wierzbicki}, \bibinfo{person}{Piotr
  Turek}, {and} \bibinfo{person}{Radoslaw Nielek}.}
  \bibinfo{year}{2010}\natexlab{}.
\newblock \showarticletitle{Learning about team collaboration from Wikipedia
  edit history}. In \bibinfo{booktitle}{{\em Proceedings of the 6th
  International Symposium on Wikis and Open Collaboration}}. ACM,
  \bibinfo{pages}{27}.
\newblock


\bibitem[\protect\citeauthoryear{Yang and Lai}{Yang and Lai}{2010}]%
        {yang2010motivations}
\bibfield{author}{\bibinfo{person}{Heng-Li Yang} {and}
  \bibinfo{person}{Cheng-Yu Lai}.} \bibinfo{year}{2010}\natexlab{}.
\newblock \showarticletitle{Motivations of Wikipedia content contributors}.
\newblock \bibinfo{journal}{{\em Computers in human behavior\/}}
  \bibinfo{volume}{26}, \bibinfo{number}{6} (\bibinfo{year}{2010}),
  \bibinfo{pages}{1377--1383}.
\newblock


\bibitem[\protect\citeauthoryear{Yu, Miao, Liu, Pan, Khalid, Shen, and
  Leung}{Yu et~al\mbox{.}}{2016}]%
        {yu2016productive}
\bibfield{author}{\bibinfo{person}{Han Yu}, \bibinfo{person}{Chunyan Miao},
  \bibinfo{person}{Siyuan Liu}, \bibinfo{person}{Zhengxiang Pan},
  \bibinfo{person}{Nur Syahidah~Bte Khalid}, \bibinfo{person}{Zhiqi Shen},
  {and} \bibinfo{person}{Cyril Leung}.} \bibinfo{year}{2016}\natexlab{}.
\newblock \showarticletitle{Productive aging through intelligent personalized
  crowdsourcing}. In \bibinfo{booktitle}{{\em 30th AAAI Conference on
  Artificial Intelligence (AAAI-16)}}.
\newblock


\end{thebibliography}

\end{document}